\documentclass{article}
\usepackage[T1]{fontenc}
\usepackage{ijcai25}

\usepackage{times}
\usepackage{soul}
\usepackage{url}
\usepackage[hidelinks]{hyperref}
\usepackage[utf8]{inputenc}
\usepackage[small]{caption}
\usepackage{graphicx}
\usepackage{subcaption}
\usepackage{amsmath}
\usepackage{amsthm}
\usepackage{amssymb}
\usepackage{booktabs}
\usepackage{tikz}
\usepackage{lipsum}
\usepackage[switch]{lineno}
\usepackage[linesnumbered,ruled,vlined]{algorithm2e}
\SetKwInOut{Parameter}{Parameter}
\newcommand{\mycommentstyle}[1]{\color[HTML]{808080}{\small #1}}
\SetKwComment{Comment}{\mycommentstyle{\# }}{}
\usepackage{multirow}
\usepackage{hyperref}
\usepackage{array}     
\usepackage{supertabular}
\usepackage{afterpage}

\title{NotaGen: Advancing Musicality in Symbolic Music Generation with \\Large Language Model Training Paradigms}

\author{
Yashan Wang$^1$\footnotemark[1] \and
Shangda Wu$^1$\footnotemark[1] \and
Jianhuai Hu$^1$\footnotemark[1] \and
Xingjian Du$^2$ \and
Yueqi Peng$^3$ \and\\
Yongxin Huang$^4$ \and
Shuai Fan$^5$ \and
Xiaobing Li$^1$ \and
Feng Yu$^1$ \and
Maosong Sun$^{1,6}$\footnotemark[2]\\
\affiliations
$^1$Central Conservatory of Music, China \and
$^2$University of Rochester, USA \and\\
$^3$Beijing Flowingtech Ltd., China \and
$^4$Independent Researcher \and\\
$^5$Beihang University, China \and
$^6$Tsinghua University, China\\
\emails
\{alexis\_wang, shangda, hujianhuai\}@mail.ccom.edu.cn,
sms@tsinghua.edu.cn\\
[1.5ex]
{\small\href{https://electricalexis.github.io/notagen-demo}{\texttt{\url{https://electricalexis.github.io/notagen-demo}}}}
}

\renewcommand{\thefootnote}{\fnsymbol{footnote}}

\begin{document}

\maketitle

\footnotetext[1]{These authors contributed equally.}
\footnotetext[2]{Corresponding author.}

\renewcommand{\thefootnote}{\arabic{footnote}}

\begin{abstract}

We introduce NotaGen, a symbolic music generation model aiming to explore the potential of producing high-quality classical sheet music. Inspired by the success of Large Language Models (LLMs), NotaGen adopts pre-training, fine-tuning, and reinforcement learning paradigms (henceforth referred to as the LLM training paradigms). It is pre-trained on 1.6M pieces of music in ABC notation, and then fine-tuned on approximately 9K high-quality classical compositions conditioned on ``period-composer-instrumentation''  prompts. For reinforcement learning, we propose the CLaMP-DPO method, which further enhances generation quality and controllability without requiring human annotations or predefined rewards. Our experiments demonstrate the efficacy of CLaMP-DPO in symbolic music generation models with different architectures and encoding schemes. Furthermore, subjective A/B tests show that NotaGen outperforms baseline models against human compositions, greatly advancing musical aesthetics in symbolic music generation.

\end{abstract}

\section{Introduction}

The pursuit of musicality is a core objective in music generation research, as it fundamentally shapes how we perceive and experience musical compositions. Symbolic music abstracts music into discrete symbols such as notes and beats, with performance signals (i.e., MIDI) and sheet music (e.g., ABC notation, MusicXML) being the two predominant modalities. Both of them can efficiently model melody, harmony, instrumentation, etc., all of which are crucial for musicality.

\begin{figure}[t] 
    \centering
    \includegraphics[width=0.5\textwidth]{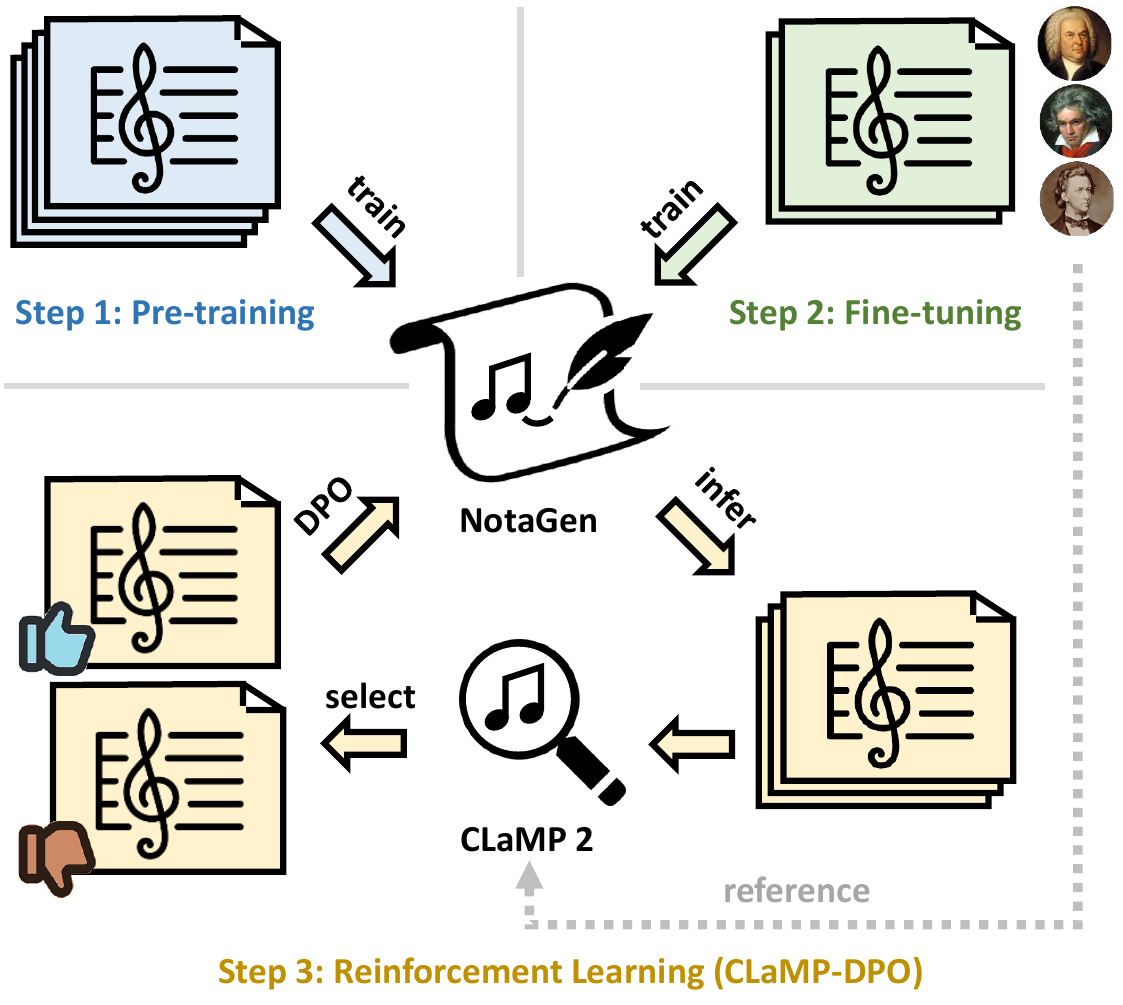} 
    \caption{An overview of NotaGen's training paradigms.}
    \label{fig:example}
\end{figure}

Training tokenized representations with language model architectures, such as Transformers \cite{vaswani2017attention}, has emerged as a powerful paradigm for symbolic music generation \cite{huang2018music,casini2022tradformer}. However, several challenges persist. First, the scarcity of high-quality music data \cite{hentschel2023annotated} hinders the ability of deep learning models to generate sophisticated compositions. Second, when optimizing a language model's loss function, the focus typically lies in minimizing the discrepancy between the predicted and the ground-truth next tokens, potentially neglecting holistic musical aspects like music structure and stylistic coherence. 

Insights from the Natural Language Processing (NLP) domain provide a promising approach to overcoming the challenges inherent in symbolic music generation. The success of Large Language Models (LLMs) \cite{dubey2024llama} has established the paradigm of pre-training, fine-tuning, and reinforcement learning as a widely acknowledged framework to improve the quality of text generation and align output with human preferences. These techniques have been successfully adapted for music generation. To overcome the scarcity of high-quality data, large-scale pre-training followed by fine-tuning on smaller, task-specific datasets has been employed effectively \cite{donahue2019lakhnes,wu2024melodyt5}. Reinforcement Learning from Human Feedback (RLHF) \cite{stiennon2020learning}, transcending next-token prediction approaches, has also shown promising results in music generation \cite{cideron2024musicrl}. However, to the best of our knowledge, the complete pipeline of LLM training paradigms has not been fully implemented in symbolic music generation. Furthermore, the high cost of RLHF for human annotation highlights the necessity for more efficient and automated solutions.

In this work, we introduce NotaGen (Musical \textbf{Nota}tion \textbf{Gen}eration), a symbolic music generation model focused on classical sheet music. Compared to MIDI generation, sheet music generation not only aims to produce artistically refined music, but also emphasizes proper voice arrangement and notation to create well-formatted sheets for performance and analysis. Furthermore, the challenge of sheet music generation is exacerbated by the diverse instrumentation and rich musicality inherent in classical music. The success of LLMs has motivated us to apply the training paradigms to sheet music generation. NotaGen is pre-trained on a corpus of over 1.6M sheets in ABC notation, and fine-tuned on a collection of approximately 9K high-quality classical pieces from 152 composers with ``period-composer-instrumentation'' (e.g.``Baroque-Bach, Johann Sebastian-Keyboard'') prompts guiding conditional generation. During reinforcement learning, we introduce the CLaMP-DPO method to further optimize NotaGen's musicality and controllability using the Direct Preference Optimization (DPO) \cite{rafailov2024direct} algorithm. In this approach, CLaMP 2 \cite{wu2024clamp}, a multimodal symbolic music information retrieval model, assigns generated samples as ``chosen'' or ``rejected'' based on references from the fine-tuning dataset. Our contributions are two-fold:

\begin{itemize}

    \item We introduce NotaGen, a symbolic music generation model implementing LLM training paradigms, which significantly outperforms baseline models in subjective A/B tests against human compositions. 
    
    \item We propose CLaMP-DPO, a reinforcement learning approach that integrates the DPO algorithm with CLaMP 2 feedback, enhancing musicality and controllability of symbolic music generation without relying on human annotation or predefined rewards. This potential is showcased across symbolic music generation models with varying architectures and encoding schemes.
    
\end{itemize}

\section{Related Works}

\subsection{Sheet Music Generation}

Sheet music generation has been widely studied, with a focus on encoding methods and composition modeling. Score Transformer \cite{suzuki2021score} introduces a tokenized representation for sheet music and applies it to piano music generation. Measure by Measure \cite{Yan-2024} models sheet music as grids of part-wise bars and employs hierarchical architectures for generation. Compared to the intricate representations used by the models above, ABC notation, a comprehensive text-based sheet music representation, simplifies encoding and facilitates composition modeling, gaining increasing adoption in recent research. The following models utilize the ABC notation: FolkRNN \cite{sturm2016music} and Tunesformer \cite{wu2023tunesformer}, specializing in folk melody generation; DeepChoir \cite{wu2023chord}, which generates choral music with chord conditioning; and MuPT \cite{qu2024mupt}, a large-scale pre-trained model for sheet music, which explores multitrack symbolic music generation. 

\subsection{Pre-training in Symbolic Music Generation}

The success of pre-training in NLP has inspired the application of this technique in symbolic music generation. LakhNES \cite{donahue2019lakhnes} enhances chiptune music generation by pre-training on the Lakh MIDI Dataset \cite{raffel2016learning}. MuseBERT \cite{wang2021musebert} adopts masked language modeling \cite{kenton2019bert}, while MelodyGLM \cite{wu2023melodyglm} implements auto-regressive blank infilling \cite{du2021glm} for generation. 
MelodyT5 \cite{wu2024melodyt5} leverages multi-task learning \cite{raffel2020exploring}. These studies highlight the effectiveness of pre-training in enhancing music generation performance.

\subsection{Reinforcement Learning in Music Generation}

Reinforcement learning has long been recognized as a promising approach for enhancing the musicality of music generation models. It has been successfully applied in RL Tuner \cite{jaques2017tuning} for melody generation, RL-Duet \cite{jiang2020rl} for online duet accompaniment, RL-Chord \cite{ji2023rl} for melody harmonization, and \cite{guo2022fine} for multi-track music generation. However, these methods either base their rewards on music theory, which limits flexibility, or tailor them to specific music styles, hindering their generalization to a broader range of music generation tasks. To tackle this problem, MusicRL \cite{cideron2024musicrl} adopts the RLHF method with extensive human feedback to align the generated compositions with human preference.

\begin{figure*}[!ht]
    \centering
    \begin{minipage}{0.49\textwidth}
        \centering
        \includegraphics[width=\textwidth]{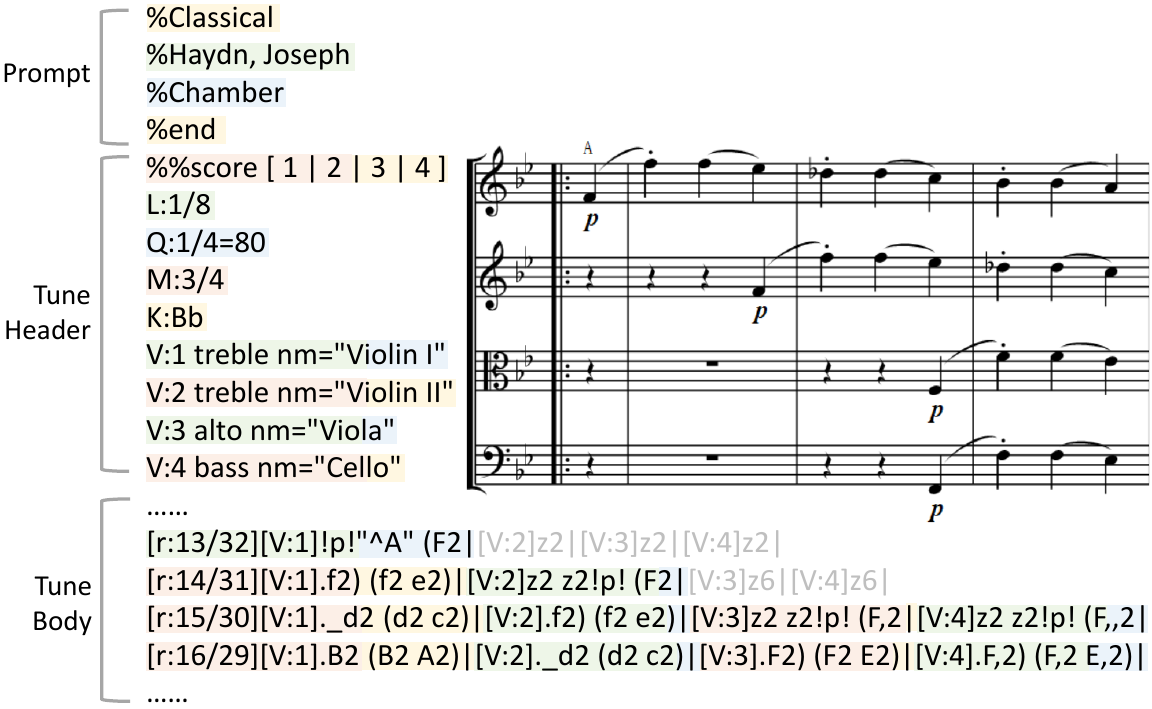}
        \caption*{(a)}
    \end{minipage}\hfill
    \begin{minipage}{0.49\textwidth}
        \centering
        \includegraphics[width=\textwidth]{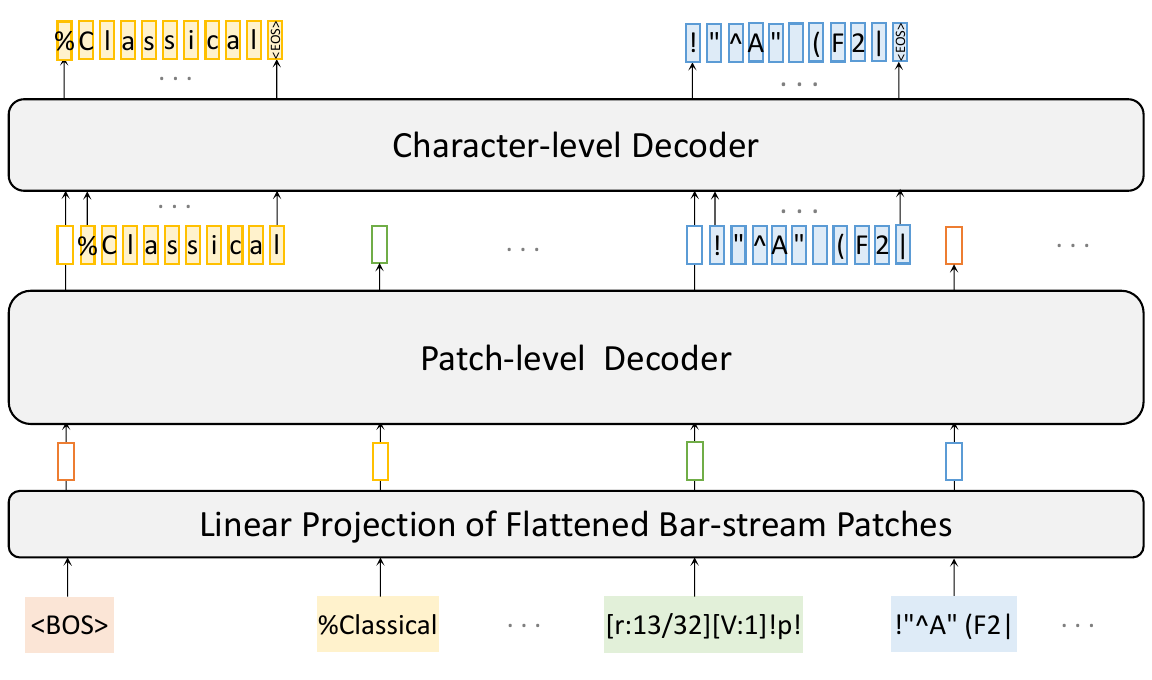}
        \caption*{(b)}
    \end{minipage}\hfill
    \caption{Data representation and model architecture of NotaGen. (a) An example of data representation for an excerpt from \textit{String Quartet in B-flat major, Hob.III:1} by Joseph Haydn using interleaved ABC notation. Bar annotations ``[r:]'' denote current/countdown bar indices, with gray bars representing omitted rests. Colored backgrounds delineate bar-stream patch boundaries. (b) The model architecture of NotaGen. After passing through the linear projection, bar-stream patches are processed by the patch-level decoder to generate features for a character-level decoder, which performs auto-regressive character prediction.}
    \label{fig:Figure 2}
\end{figure*}

\section{NotaGen}

\subsection{Data Representation}

ABC notation sheets consist of two parts: the tune header, which contains metadata such as tempo, time signature, key, and instrumentation; the tune body, where the musical content for each voice is recorded. We adopt a modified version---interleaved ABC notation \cite{wu2024clamp,qu2024mupt}. In this format, different voices of the same bar are rearranged into a single line and differentiated using voice indicators ``[V:]''. This ensures alignment of duration and musical content across voices. Furthermore, we remove bars with full rests (containing only ``z'' or ``x'' notes), reducing the length to 80.7\% on average, while increasing information density.

We employ stream-based training and inference methods to enable long musical piece generation. We annotate the current and countdown bar indices before each tune body line using the label ``[r:]''. During training, we randomly segment the tune body and concatenated it with the tune header for longer pieces; during inference, we enforce the generation to start from scratch using the bar annotations. If the piece is incomplete within the current context length, we concatenate the generated tune header with the second half of the tune body and continue generating until the final bar.

\subsection{Model Architecture}

NotaGen utilizes bar-stream patching \cite{wang2024exploring} and the Tunesformer architecture \cite{wu2023tunesformer}. Building upon bar patching \cite{wu2023clamp}, bar-stream patching divides the tune header lines and bars into fixed-length patches (padded when necessary), striking a balance between musicality of generation and computational efficiency among sheet music tokenization methods. 

NotaGen consists of two hierarchical GPT-2 decoders \cite{radford2019language}: a patch-level decoder and a character-level decoder. Each patch is flattened by concatenating one-hot character vectors and then passed through a linear layer to obtain the patch embedding. The patch-level decoder captures the temporal relationships among patches, and its final hidden states are passed to the character-level decoder, which auto-regressively predicts the characters of the next patch. The data representation and model architecture are illustrated in Figure \ref{fig:Figure 2}.

\subsection{Training Paradigms}

\subsubsection{Pre-training}

Pre-training enables NotaGen to capture fundamental musical structures and patterns through next-token prediction on a large, diverse dataset spanning various genres and instrumentations.

The pre-training stage utilized a carefully curated internal-use dataset comprising 1.6M ABC notation sheets. We also preprocessed the text annotations, retaining music-related content such as tempo and expression hints, while removing irrelevant content like lyrics and background information.

All music sheets were transposed to 15 keys (including F$\sharp$, G$\flat$, C$\sharp$, C$\flat$) for data augmentation. During training, a randomly selected key was used for each piece in every epoch.

\subsubsection{Fine-tuning}

NotaGen was fine-tuned on high-quality classical music sheet data to further enhance musicality in generation. Spanning from the intricate contrapuntal orchestra suites of the Baroque period to the melodious and harmonically nuanced piano pieces of the Romantic era, classical music encompasses a diverse array of compositional styles and instrumentations, all characterized by exceptional musicality.

Thus, we curated a fine-tuning dataset comprising 8,948 classical music sheets, from DCML corpora \cite{neuwirth2018annotated,hentschel2021semi,hentschel2021annotated,hentschel2023annotated}, OpenScore String Quartet Corpus \cite{gotham2023openscore}, OpenScore Lieder Corpus \cite{gotham2022openscore}, ATEPP \cite{zhang2022atepp}, KernScores \cite{sapp2005online}, and internal resources, as listed in Table \ref{tab:FinetuneDataset}. Sheets with more than 16 staves were excluded due to generation complexity. Each work was assigned with three labels: period, composer and instrumentation. The data distribution is provided in supplementary materials, and the details of each label are explained as follows:

\begin{itemize}
    
    \item \textbf{Period}: 
    \begin{itemize}
        \item \textbf{Baroque} (1600s-1750s): e.g., Bach, Vivaldi.
        \item \textbf{Classical} (1750s-1810s): e.g., Mozart, Beethoven.
        \item \textbf{Romantic} (1810s-1950s): e.g., Chopin, Liszt. 
    \end{itemize}

    \item \textbf{Composer}: The official names of a total of 152 composers, as listed on IMSLP\footnote{\url{https://imslp.org/}}, were included.

    \item \textbf{Instrumentation}: 
    \begin{itemize}
        \item \textbf{Keyboard}: piano and organ works.
        \item \textbf{Chamber}: instrumental music typically for a small group of performers, each playing a unique part.
        \item \textbf{Orchestral}: instrumental music for orchestra. 
        \item \textbf{Art Song}: vocal music typically for solo or duet voices with piano accompaniment. 
        \item \textbf{Choral}: vocal music for a choir.
        \item \textbf{Vocal-Orchestral}: works involving both vocal and orchestral elements, including Cantata, Oratorio, and Opera.
    \end{itemize}
    
\end{itemize}

\begin{table}[t]
    \centering
    \begin{tabular}{lr}
        \toprule
        Data Sources  & Amount \\
        \midrule
        \textit{DCML Corpora}                       & 560 \\
        \textit{OpenScore String Quartet Corpus}    & 342 \\
        \textit{OpenScore Lieder Corpus}            & 1,334 \\
        \textit{ATEPP}                              & 55 \\
        \textit{KernScores}                         & 221 \\
        \textit{Internal Sources}                   & 6,436 \\
        \midrule
        \textit{Total}                              & 8,948 \\
        \bottomrule
    \end{tabular}
    \caption{Data sources and the respective amounts for fine-tuning.}
    \label{tab:FinetuneDataset}
\end{table}

In fine-tuning, a ``period-composer-instrumentation'' prompt was prepended to each piece for conditional generation. This approach challenges NotaGen to produce high-quality compositions, imitate the styles of composers across different periods, and conform to specified instrumentation requirements.

To facilitate NotaGen's learning of appropriate pitch ranges for each instrument while optimizing data utilization, data augmentation during fine-tuning was restricted to the six nearest key transpositions of the original. Keys farther from the original were selected with decreasing probability.

\subsubsection{Reinforcement Learning}

To refine both the musicality and the prompt controllability of the fine-tuned NotaGen, we present CLaMP-DPO. This method builds upon the principles of Reinforcement Learning from AI Feedback (RLAIF) \cite{DBLP:conf/icml/0001PMMFLBHCRP24} and implements Direct Preference Optimization (DPO) \cite{rafailov2024direct}.  In CLaMP-DPO, CLaMP 2 serves as the evaluator within the DPO framework, distinguishing between chosen and rejected musical outputs to optimize NotaGen.

CLaMP 2 is a multimodal symbolic music information retrieval model supporting both ABC notation and MIDI formats. Leveraging contrastive learning, CLaMP 2 extracts semantic features that encapsulate global musical properties. These features encompass comprehensive musical information, including style, instrumentation, and compositional complexity. Meanwhile, they are consistent with human subjective perceptions, as validated by \cite{retkowski2024frechet}. In the context of music generation, the objective is to produce pieces which closely resemble the ground truth. Accordingly, it is critical to ensure the alignment of the semantic features between the generated pieces and authentic references.

We introduce the CLaMP 2 Score to quantify the similarity among pieces. To elaborate, we denote $P$ as the set of prompts for NotaGen. For each prompt $p \in P$, $Y_p$ represents the corresponding set of ground truth with an average semantic feature $\bar{z}_{p}$. Similarly, each prompt $p$ has a generated set $X_p$, where each piece $x_p$ is associated with a semantic feature $z_{x_{p}}$.

The CLaMP 2 score $c$ for a generated piece $x_{p}$ is defined as the cosine similarity between $z_{x_{p}}$ and $\bar{z}_{p}$:

\begin{equation}
    c_{x_{p}} = \frac{z_{x_{p}} \cdot \bar{z_p}}{\| z_{x_{p}} \| \| \bar{z_p} \|}.
\label{clamp2score}
\end{equation}

Our goal is to maximize the average, $\bar{c}_{x_p}$ over $X_p$, thereby ensuring the music generated for prompt $p$ aligns semantically with the ground truth. It is achieved by employing the DPO algorithm to improve $\bar{c}_{x_p}$. 


The DPO algorithm optimizes a language model based on preference data, which consists of paired chosen and rejected examples under the same prompts. It eliminates the need of explicit reward modeling. In the proposed CLaMP-DPO algorithm, the fine-tuned model first generates data across the prompt set $P$. For each generated set $X_p$, the pieces $x_{p} \in X_p$ are sorted according to $c_{x_{p}}$, with the top 10\% selected as chosen set $X_{pw}$ and the bottom 10\% as rejected set $X_{pl}$. Additional criteria, such as syntax error checks or the exclusion of ground-truth plagiarism, can be applied to refine these sets. Finally, the chosen and rejected pairs $(x_{pw},x_{pl})$ are randomly selected and combined into preference data for optimization.

Given a prompt $p$, an auto-regressive language model predicts the next token based on its policy $\pi_\theta$, where $\theta$ represents the model parameters. The probability of generating a chosen data $x_{pw}$ is $\pi_\theta(x_{pw}|p)$, and that of generating a rejected data $x_{pl}$ is $\pi_\theta(x_{pl}|p)$. To prevent excessive drift from the initial model that generates the preference data and ensure diversity in the generated content, the initial model policy, i.e., the reference model policy $\pi_{\text{ref}}$ is introduced and kept frozen during optimization. The objective function to be minimized in DPO is given by:

\begin{algorithm}[tb]
    \SetAlgoLined
    \caption{Iterative CLaMP-DPO}
    \label{alg:clamp_dpo}
    \KwIn{
        Fine-tuned policy $\pi_\theta^0$, 
        CLaMP 2 model $C$, 
        prompt set $P$, 
        fine-tuning dataset $Y$}
    \Parameter{
        Iterations $K$, 
        DPO hyperparameter $\beta$, 
        DPOP hyperparameter $\lambda$, 
        optimization steps $N$, 
        learning rate $\eta$}
    \KwOut{Optimized policy $\pi_\theta^K$}
    
    \Comment{\textcolor{gray}{Initialize ground-truth features}}
    \ForEach{\textnormal{prompt} $p \in P$}{
        $\bar{z}_p \gets \text{Avg}(C(y_p)), \quad \forall y_p \in Y_p$
    }
    \Comment{\textcolor{gray}{Iterative Optimization}}
    \For{$k \gets 1$ \KwTo $K$}{
        \Comment{\textcolor{gray}{Construct preference data}}
        \ForEach{\textnormal{prompt} $p \in P$}{
            $X_p^{k-1} \gets \pi_\theta^{k-1}(p)$ \Comment{\textcolor{gray}{Generate on p}}
            \ForEach{\textnormal{piece} $x_{p}^{k-1} \in X_P^{k-1}$}{
                ${z}_{x_p^{k-1}} \gets C(x_{p}^{k-1})$ \\
                $c_{x_p^{k-1}} \gets \text{Eq.~\eqref{clamp2score}}(z_{x_p^{k-1}}, \bar{z}_p)$
            }
            $X_{pw}^{k-1}, X_{pl}^{k-1} \gets \text{Select}(X_p^{k-1}, \text{Sort}(c_{x_p^{k-1}}))$ 
        }
        \Comment{\textcolor{gray}{Optimize using DPO}}
        $\pi_\text{ref} \gets \pi_\theta^{k-1}$ \\
        \For{$i \gets 1$ \KwTo $N$}{
            Sample prompt $p \sim P$ \\
            Sample pairs $(x_{pw}, x_{pl}) \sim (X_{pw}^{k-1}, X_{pl}^{k-1})$ \\
            $\theta \gets \theta - \eta \nabla_\theta \mathcal{L}_\text{DPOP}(\pi_\theta, \pi_\text{ref}, x_{pw}, x_{pl}, p, \beta, \lambda)$
        }
        $\pi_\theta^k \gets \pi_\theta$ \\
    }
    \KwRet{$\pi_\theta^K$}
\end{algorithm}

\begin{equation}
\begin{aligned}
    \mathcal{L}_{\text{DPO}}(\pi_\theta; \pi_{\text{ref}}) &= - \mathbb{E}_{(p, x_{pw}, x_{pl}) \sim \mathcal{D}} \Big[ \log \sigma \Big( \beta \log \frac{\pi_\theta(x_{pw} | p)}{\pi_{\text{ref}}(x_{pw} | p)} \\
    &\quad - \beta \log \frac{\pi_\theta(x_{pl} | p)}{\pi_{\text{ref}}(x_{pl} | p)} \Big) \Big],
\end{aligned}
\label{DPO}
\end{equation}

\noindent where $\sigma$ is the sigmoid function, $\mathcal{D}$ is the preference dataset, and $\beta$ is the hyperparameter that controls the deviation between $\pi_{\theta}$ and $\pi_{\text{ref}}$.

The optimization process increases the relative log probability of chosen data over rejected data. However, we observed a decrease in $\pi_\theta(x_{pw}|p)$, leading to degraded outputs. To mitigate this issue, we adopt the DPO-Positive (DPOP) objective function \cite{pal2024smaug}, which incorporates a penalty term to stabilize $\pi_\theta(x_{pw}|p)$:

\begin{equation}
\begin{aligned}
    \mathcal{L}_{\text{DPOP}}(\pi_\theta; \pi_{\text{ref}}) &= - \mathbb{E}_{(p, x_{pw}, x_{pl}) \sim \mathcal{D}} \Big[ \log \sigma \Big( \beta \log \frac{\pi_\theta(x_{pw} | p)}{\pi_{\text{ref}}(x_{pw} | p)} \\
    &\quad - \beta \log \frac{\pi_\theta(x_{pl} | p)}{\pi_{\text{ref}}(x_{pl} | p)} \\
    &\quad - \beta \lambda \cdot \text{max} \Big(0, \log \frac{\pi_\text{ref}(x_{pw} | p)}{\pi_{\theta}(x_{pw} | p)} \Big) \Big) \Big],
\end{aligned}
\label{DPOP}
\end{equation}

\noindent where the hyperparameter $\lambda$ controls the impact of penalty.

The fine-tuned model is optimized by minimizing $\mathcal{L}_{\text{DPOP}}$ in Eq.\eqref{DPOP} for a specified number of steps, completing the process of CLaMP-DPO algorithm. Notably, CLaMP-DPO supports iterative optimization. After the first round, the model generates a new set $X_p^\prime$. Using CLaMP 2, we construct new chosen and rejected sets, $X_{pl}^\prime$ and $X_{pw}^\prime$, allowing the model to undergo further optimization via Eq.\eqref{DPOP}.

\section{Experiments}

\subsection{Settings}

The experiments are divided into two parts. The first part assesses CLaMP-DPO's ability to improve the controllability and musicality of symbolic music models. The second part compares the musicality of NotaGen with baseline models. Along with the pre-trained NotaGen, we selected two pre-trained symbolic music generation models as baselines: MuPT \cite{qu2024mupt} and Music Event Transformer (MET) \footnote{\url{https://huggingface.co/skytnt/midi-model-tv2o-medium}}\cite{skytnt2024midimodel}. All models adopt language model architectures and are trained auto-regressively. A brief overview of their architectures and pre-training procedures follows:

\begin{itemize}

    \item \textbf{NotaGen} features a 20-layer patch-level decoder and a 6-layer character-level decoder, with a context length of 1024 and a hidden size of 1280, totaling 516M parameters. It was pre-trained on 1.6M ABC notation sheets, augmented to 15 key transpositions. The AdamW optimizer \cite{loshchilov2019decoupled} was utilized with a learning rate of 1e-4 and a 1,000-step  warm-up phase. The pre-training was performed on 8 NVIDIA H800 GPUs, with a batch size of 4 per GPU.

    \item \textbf{MuPT} utilizes Synchronized Multi-Track ABC notation (SMT-ABC) as data representation. SMT-ABC is equivalent to interleaved ABC notation, as both merge multi-track voices into a single sequence. Byte Pair Encoding (BPE) is used for tokenization. MuPT-v1-8192-550M, the chosen baseline model, consists of a 16-layer Transformer decoder with a hidden size of 1024 and a context length of 8192, totaling 505M parameters. MuPT was pre-trained on a corpus of 33.6B tokens.

    \item \textbf{MET} encodes MIDI events into token sequences and uses hierarchical Transformer decoders for generation, including a event-level decoder and a token-level decoder. Details on the encoding and model architecture are provided in the supplementary materials. MET consists of a 12-layer event-level decoder and a 3-layer token-level decoder, with a context length of 4096 and a hidden size of 1024, totaling 234M parameters. It was pre-trained on three MIDI datasets: Los Angeles MIDI Dataset \cite{lev2024losangelesmididataset}, Monster MIDI Dataset \footnote{\url{https://huggingface.co/datasets/projectlosangeles/Monster-MIDI-Dataset}}, and SymphonyNet Dataset \cite{liu2022symphony}.
    
\end{itemize}

We applied fine-tuning and reinforcement learning to these models using their pre-trained weights.
    
\paragraph{Fine-tuning.} The fine-tuning dataset for NotaGen and MuPT comprises 8,948 classical music pieces, referred to as the sheet ground truth set (sheet-GT). All data were formatted to match the pre-training data of different models, each preceded by a ``period-composer-instrumentation'' prompt. Due to the challenges in converting between MIDI and ABC notation, only the keyboard subset, consisting of 3,104 pieces was used for fine-tuning MET, referred to as the MIDI ground truth set (MIDI-GT). Each piece was preceded by a ``period-composer'' prompt.

\paragraph{Reinforcement learning.} Considering that the accuracy of prompt semantic feature $\bar{z}_p$ in CLaMP-DPO relies on a sufficient amount of ground truth data in $Y_p$, we defined the prompt set $P$ to only include prompts $p$ that appear more than ten times in the fine-tuning dataset ($Y_p > 10$). The detailed list of $P$ can be referred in supplementary materials. For NotaGen and MuPT, $P$ contained 112 prompts, covering 86.4\% of the data; for MET, $P$ contained 29 prompts, covering 90.5\%. The number of iterations $K$ was set to 3, with approximately 100 pieces generated per prompt as $X_p$ in each iteration. The chosen and rejected sets were constructed based on CLaMP 2 Scores. Sheets where staves for the same instrument were not grouped together were excluded from the chosen set. The hyperparameters $\beta=0.1$ and $\lambda=10$ were used, with $N=10,000$ optimization steps. The learning rate was fixed at 1e-6 for NotaGen and MET, and 1e-7 for MuPT, yielding stable CLaMP-DPO performance.

Given the challenge of establishing objective metrics that fully capture musicality, we conducted subjective A/B tests in both experiments to evaluate different models and settings. For each question, two pieces were generated using identical prompts; videos were rendered from sheet music using Sibelius and MIDI files using MIDIVisualizer\footnote{\url{https://github.com/kosua20/MIDIVisualizer}}. Participants were instructed to evaluate musicality from multiple perspectives and select the piece they found more musically appealing, or indicate no preference if they perceived no differences. The evaluation criteria included melodic appeal, harmonic fluency, orchestral balance, counterpoint correctness, and structural coherence, and, for sheet music, notation formatting quality. A total of 92 participants from music colleges took part in the assessment. At least 35 valid responses were recorded for each test group, ensuring statistical reliability.

\subsection{Ablation Studies on CLaMP-DPO}

This experiment evaluates the impact of the proposed CLaMP-DPO algorithm in enhancing the controllability and musicality of generated music for NotaGen, MuPT, and MET. In the objective assessment, we selected several metrics for both the fine-tuned models (denoted as $K=0$) and the models after $K$ iterations of CLaMP-DPO optimization. We also assessed a subset of these metrics on the fine-tuning datasets (sheet-GT and MIDI-GT) for reference. The metrics are as follows:

\begin{itemize}

    \item \textbf{Average CLaMP 2 Score (ACS)}: The average CLaMP 2 Score across generated outputs. For sheet-GT and MIDI-GT, the score is computed over the corpus.

    \item \textbf{Label Accuracy (LA)}: The alignment with specified period (per.) and instrumentation (inst.) prompts. We extracted features from the fine-tuning dataset via a multimodal symbolic music encoder---M3\cite{wu2024clamp}, then trained two linear classifiers to predict the period and instrumentation labels. LA is defined as the classification accuracy, where for the fine-tuning dataset, it measures the accuracy on the test set, and for generated outputs, it reflects the match between predicted labels and prompt labels.

    \item \textbf{Bar Alignment Error (BAE)}: The proportion of bars where duration is misaligned, occurring in either the generated output or the fine-tuning corpus. This metric applies only to sheet data.
    
    \item \textbf{Perplexity (PPL)}: A language model metric, where lower PPL indicates better prediction capability.
    
\end{itemize}

\begin{table}[!h]
    \centering
    \resizebox{0.5\textwidth}{!}{ 
    \begin{tabular}{l|c|ccccc}
    \toprule
    \multirow{2}{*}{\textbf{Models \& Data}} & \multirow{2}{*}{\textbf{\textit{K}}} & \multirow{2}{*}{\textbf{ACS}} & \multicolumn{2}{c}{\textbf{LA (\%)}} & \multirow{2}{*}{\textbf{BAE (\%)}} & \multirow{2}{*}{\textbf{PPL}}   \\ 
    \cmidrule{4-5}
                                    &           &               & \textbf{Per.}     & \textbf{Inst.}    &                   &           \\ 
    \midrule
    sheet-GT                        & -     & 0.792             & 96.1              & 95.5              & 0.377             & -                                    \\
    \midrule
    \multirow{4}{*}{NotaGen}        & 0     & 0.570             & 84.7              & 78.5              & 0.269             & \textbf{1.2151}                \\
                                    & 1     & 0.674             & 92.1              & 87.8              & 0.175             & 1.2341                         \\
                                    & 2     & 0.708             & \underline{\textbf{93.3}}     & 92.9              & \underline{\textbf{0.158}}    & 1.2614                \\
                                    & 3     & \textbf{0.730}    & 93.0              & \underline{\textbf{94.6}}     & 0.176             & 1.2880                         \\
    \midrule
    \multirow{4}{*}{MuPT}           & 0     & 0.515             & 76.3              & 78.6              & \textbf{0.824}    & \textbf{1.4159}       \\
                                    & 1     & 0.596             & 78.8              & 86.2              & 1.520             & 1.4476                         \\
                                    & 2     & 0.631             & 80.3              & \textbf{89.2}     & 2.601             & 1.5214                         \\
                                    & 3     & \textbf{0.674}    & \textbf{82.1}     & 87.6              & 4.676             & 1.6121                         \\
    \midrule
    \midrule
    MIDI-GT                         & -     & 0.812             & 92.9              & -                 & -                 & -                 \\
    \midrule
    \multirow{4}{*}{MET}            & 0     & 0.565             & 30.0              & -                 & -                 & \textbf{1.2251}                    \\
                                    & 1     & 0.609             & 34.6              & -                 & -                 & 1.2261                             \\
                                    & 2     & 0.637             & 36.7              & -                 & -                 & 1.2255                             \\
                                    & 3     & \textbf{0.655}    & \textbf{38.2}     & -                 & -                 & 1.2290                             \\
    \bottomrule
    \end{tabular}
    }
    \caption{Objective metrics on fine-tuned models and the models after each iteration of CLaMP-DPO optimization. Some of metrics were also assessed on the fine-tuning dataset for reference.}
    \label{tab:Obejctive clampdpo}
\end{table}

\begin{figure}[!t] 
    \centering
    \includegraphics[width=0.5\textwidth]{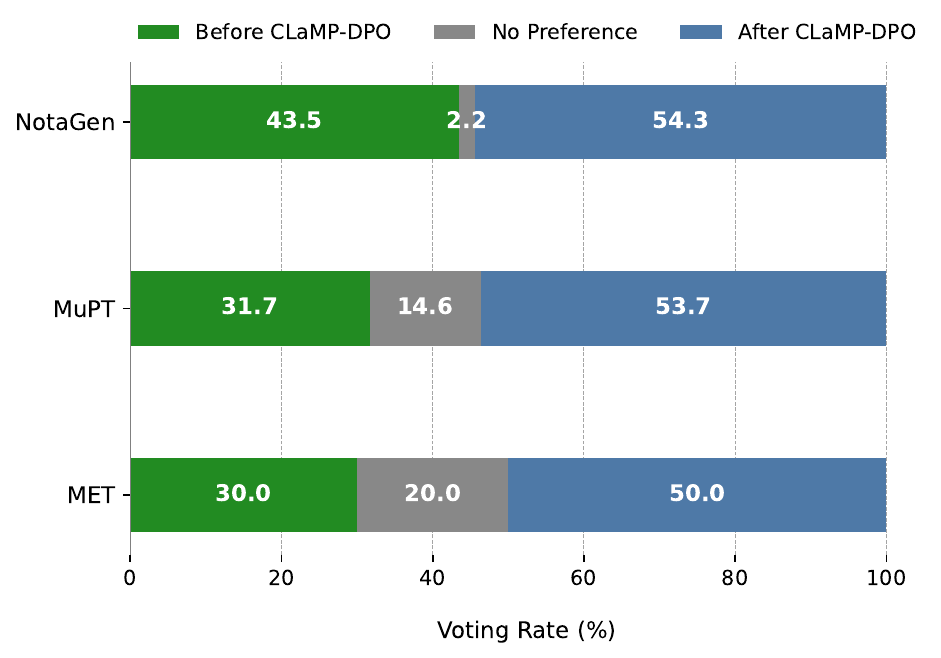} 
    \caption{Subjective A/B tests on musicality of generated outputs before and after CLaMP-DPO optimization. All models exhibited improvement in human-perveiced musicality after applying the CLaMP-DPO algorithm.}
    \label{fig:Subjective clampdpo}
\end{figure}

We conducted subjective A/B tests on each model before and after three optimization iterations with CLaMP-DPO to appraise its efficacy in enhancing the musical quality of generated outputs. The results of the objective and subjective tests are presented in Table \ref{tab:Obejctive clampdpo} and Figure \ref{fig:Subjective clampdpo}, respectively. 

The ACS, as the primary optimization goal, exhibited a monotonic increase across iterations of its DPO-based process. Though significant improvements were observed in early iterations, subsequent gains exhibited diminishing returns.

LA for period and instrumentation classification exceeded 90\% on the test set of fine-tuning data, validating the reliability of the label assignments and the performance of the classifiers. Following the CLaMP-DPO method, all models demonstrated a noticeable improvement in LA, indicating enhanced prompt controllability and better alignment with the intended musical styles. NotaGen exhibited the highest controllability among the models, further confirming its superior adaptability to specified prompts.

Regarding BAE, NotaGen maintained a relatively low error rate throughout optimization, indicating its character-level prediction is more robust at managing duration consistency. In contrast, MuPT's increased error rate is likely due to the use of BPE tokenization, which may merge duration with other musical elements, such as pitch, into single tokens. It may lead to inaccuracies in duration prediction after CLaMP-DPO adjusts token probabilities. 

Subjective A/B tests showed that all models exhibited improvement in musicality after applying the CLaMP-DPO algorithm, with post-optimization outputs receiving more votes than their pre-optimization counterparts. However, it is noteworthy that PPL increased after optimization. It suggests that PPL may not be a suitable indicator for model performance in symbolic music generation, highlighting the limitations of traditional language model metrics in assessing musical quality.

In summary, the CLaMP-DPO algorithm efficiently enhanced both the controllability and the musicality across three models, irrespective of their data modalities, encoding schemes, or model architectures. This underscores CLaMP-DPO's broad applicability and potential for auto-regressively trained symbolic music generation models.

\subsection{Comparative Evaluations}

\begin{figure}[t] 
    \centering
    \includegraphics[width=0.5\textwidth]{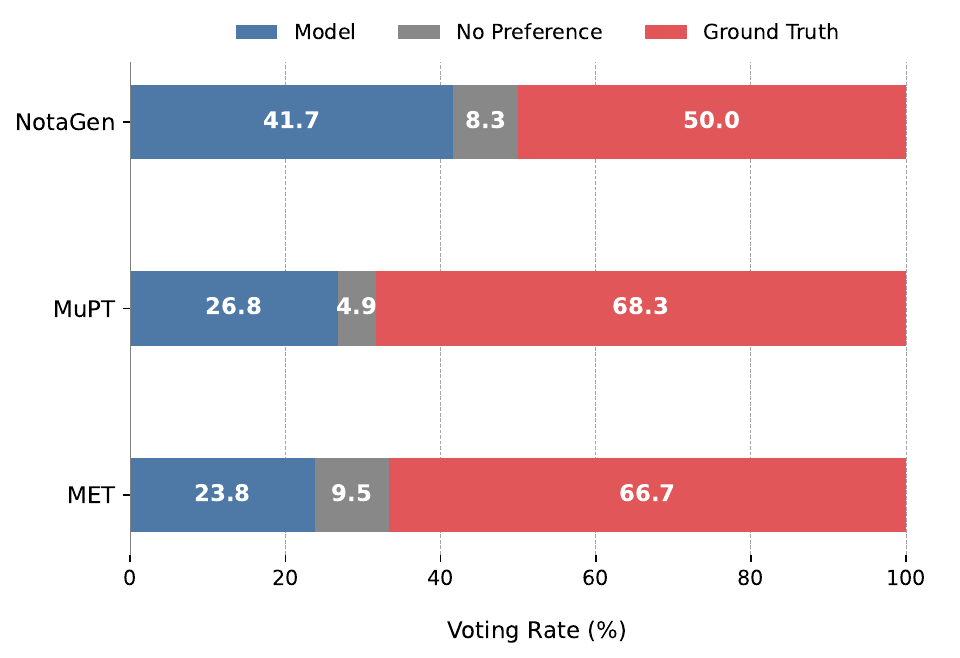} 
    \caption{Subjective A/B test between model outputs and ground truth. NotaGen achieved the highest voting rate against the ground truth among the three models.}
    \label{fig:Comparative}
\end{figure}

This experiment compares the musicality of three models after the LLM training paradigm. For baseline comparison, we constructed the reference set using human-authored musical pieces from the fine-tuning dataset, which represent professional compositional standards. The subjective A/B tests were organized into three groups, each containing the generated results of a model and the ground truth. For comparison involving MET, all data were converted to MIDI to eliminate format-based bias. The results are shown in Figure \ref{fig:Comparative}.

Human compositions consistently outperformed all model-generated outputs in voting due to their exceptional musicality. Nevertheless, NotaGen achieved the highest voting rate against the ground truth among the three models, suggesting its superior perceived musicality relative to other systems in human evaluations.

Overall, NotaGen outperformed the baseline models. The superior performance of NotaGen compared to MuPT is attributed to well-designed data representation and tokenization. Despite its architectural similarities to MET, NotaGen achieved better musicality, benefiting from the efficiency and structural integrity of sheet music representation compared to MIDI.

\section{Limitations and Challenges}

While NotaGen shows promising advancements in symbolic music generation, limitations and challenges still warrant discussion.

We once introduced a post-training stage between pre-training and fine-tuning, refining the model with classical-style subset of the pre-training dataset. While it accelerated the fine-tuning convergence and improved ACS for NotaGen, the impact was less pronounced on MuPT and MET.

Furthermore, the prerequisite for evaluating generated results using CLaMP 2 Score is that the model has been well trained and is capable of generating reasonable compositions. For corrupted or syntactically flawed pieces, the CLaMP 2 Score may not reliably indicate the true musical similarity.

Finally, we found that modeling orchestral music presents greater challenges compared to smaller ensembles (e.g. solo piano or string quartets). While rest-bar omission during data pre-processing addresses the degeneration due to excessive blank bars in ensemble writing, NotaGen's performance in orchestral music generation still lags behind. More effective methods are expected for generating large ensemble compositions.

\section{Conclusions}

In this work, we present NotaGen, a symbolic music generation model designed to advance the musicality of generated outputs through a comprehensive LLM-inspired training paradigm. By integrating pre-training, fine-tuning, and reinforcement learning with the proposed CLaMP-DPO algorithm, NotaGen demonstrates superior performance in generating compositions that align with both the music style specified by prompts and human-perceived musicality. Our experiments validate two key findings: (1) CLaMP-DPO efficaciously enhances controllability and musicality across diverse symbolic music models, regardless of their modality, architectures, or encoding schemes, without requiring human annotations or predefined rewards; (2) NotaGen outperforms baseline models in subjective evaluations, achieving the highest voting rate against human-composed ground truth.

NotaGen establishes the viability of adapting LLM training paradigms to symbolic music generation, while addressing domain-specific challenges, including data scarcity and demand for high-quality music outputs. Future work could extend this framework with CLaMP-DPO to broader musical genres such as jazz, pop, and ethnic music; as well as exploring its compatibility with emerging music generation models.

\newpage

\section*{Acknowledgments}

We would like to express our sincere gratitude to SkyTNT, the author of MET, for his valuable support on this project. We also acknowledge Yuling Yang, Xinran Zhang, Jiafeng Liu, Yuqing Cheng, and Yuhao Ding from Central Conservatory of Music for their support, especially on subjective tests and paper writing. 

This work was supported by the following funding sources: Special Program of National Natural Science Foundation of China (Grant No. T2341003), Advanced
Discipline Construction Project of Beijing Universities, Major Program of National Social Science Fund of China (Grant No. 21ZD19), and the National Culture and Tourism Technological Innovation Engineering Project (Research and Application of 3D Music).

\bibliographystyle{named}
\bibliography{ijcai25}

\newpage


\maketitle

\appendix

\section{MIDI Event Transformer}

MIDI Event Transformer (MET) is a pre-trained music generation model. It encodes MIDI events into token sequences and utilizes hierarchical transformer decoders for generation. 

\subsection{MIDI Encoding}

In the MIDI encoding process, each MIDI event is transformed into a token sequence $s_i$. This sequence begins with the token representing the event type $e_i$, followed by tokens corresponding to the event parameters $p_i^j$, as shown below:

\begin{equation}
s_i = \{e_i, p_i^1, p_i^2, ..., p_i^n\}
\end{equation}

The types of events and their corresponding parameters are listed in Table~\ref{tab:event list}. Among these, BOS (beginning-of-sequence) and EOS (end-of-sequence) events are used to mark the start and end of a musical piece, respectively. Parameters such as channel (16 values), pitch (128 values), velocity (128 values), controller (128 values), program (128 values), controller value (128 values) all comply with the MIDI standard.  The details of additional parameters are as follows:

\begin{itemize}

\item Time 1 and time 2: Represent the timing of an event. The absolute time position is calculated as the cumulative sum of ticks preceding the event, and the beat position is obtained by dividing this sum by the ticks per beat. Time 1 captures the beat difference between the current and previous events, with a vocabulary size of 128. To enable finer-grained positioning, a beat is subdivided into 16 parts, with time 2 indicating which subdivision the event falls into.

\item Track: Describes the track in which the event occurs, with a maximum of 128 possible tracks.

\item Duration: The note duration is calculated using a resolution of 64th notes, with an upper limit of 2048, which allows for the representation of notes up to 128 beats long.

\item BPM: Beats per minute, with a range from 1 to 384.

\item Numerator and denominator: The numerator can range from 1 to 16, while the denominator can be one of 2, 4, 8 or 16.

\item Key signature accidentals: Indicates the number of sharps or flats in the key signature. The value ranges from -7 (7 flats, C$\flat$) to 7 (7 sharps, C$\sharp$).

\item Mode: Specifies whether the key is major or minor, with values between 0 and 1.

\end{itemize}

Additionally, as a special token type for $p_i^j$, \texttt{<PAD>} is used to pad all sequences to a uniform length of 8. This results in a total vocabulary size of 3406 unique tokens.

During the fine-tuning and reinforcement learning stages of our experiments, we extend the original vocabulary by including 3 period IDs and 36 composer IDs. These are prepended as "period" and "composer" events—each represented by a sequence of 8 identical IDs—prior to the encoded MIDI sequences, serving as prompts.

\begin{table}[t]
    \centering
    \begin{tabular}{ll}
        \toprule
        Event  & Parameters \\
        \midrule
        \multirow{2}{*}{Note}           & time 1, time 2, track, channel, pitch,  \\
                                        & velocity, duration \\
        \multirow{2}{*}{Program Change} & time 1, time 2, track, channel, \\
                                        & program \\
        \multirow{2}{*}{Control Change} & time 1, time 2, track, channel,  \\
                                        & controller, controller value \\
        Set Tempo                       & time 1, time 2, track, bpm \\
        \multirow{2}{*}{Time Signature} & time 1, time 2, track, numerator,  \\
                                        & denominator \\
        \multirow{2}{*}{Key Signature}  & time 1, time 2, track, key signature \\
                                        & accidentals, mode \\
        BOS                             & \texttt{<BOS>} \\
        EOS                             & \texttt{<EOS>} \\
        \bottomrule
    \end{tabular}
    \caption{Event and parameter types in MIDI Event Transformer's encoding.}
    \label{tab:event list}
\end{table}

\subsection{Model Architecture}

The MIDI Event Transformer is comprised of two hierarchical decoders: an event-level decoder and a token-level decoder, both of which are based on the Llama architecture.

Initially, each event token sequence is processed through a token embedding layer, where individual token embeddings are aggregated to produce a dense vector representation. The event-level decoder focuses on modeling temporal dependencies across high-level events, thereby capturing their sequential relationships. The output hidden states from the event-level decoder are subsequently passed into the token-level decoder, which generates the detailed token sequence in an auto-regressive manner.

\begin{figure}[t] 
\centering
\includegraphics[width=0.5\textwidth]{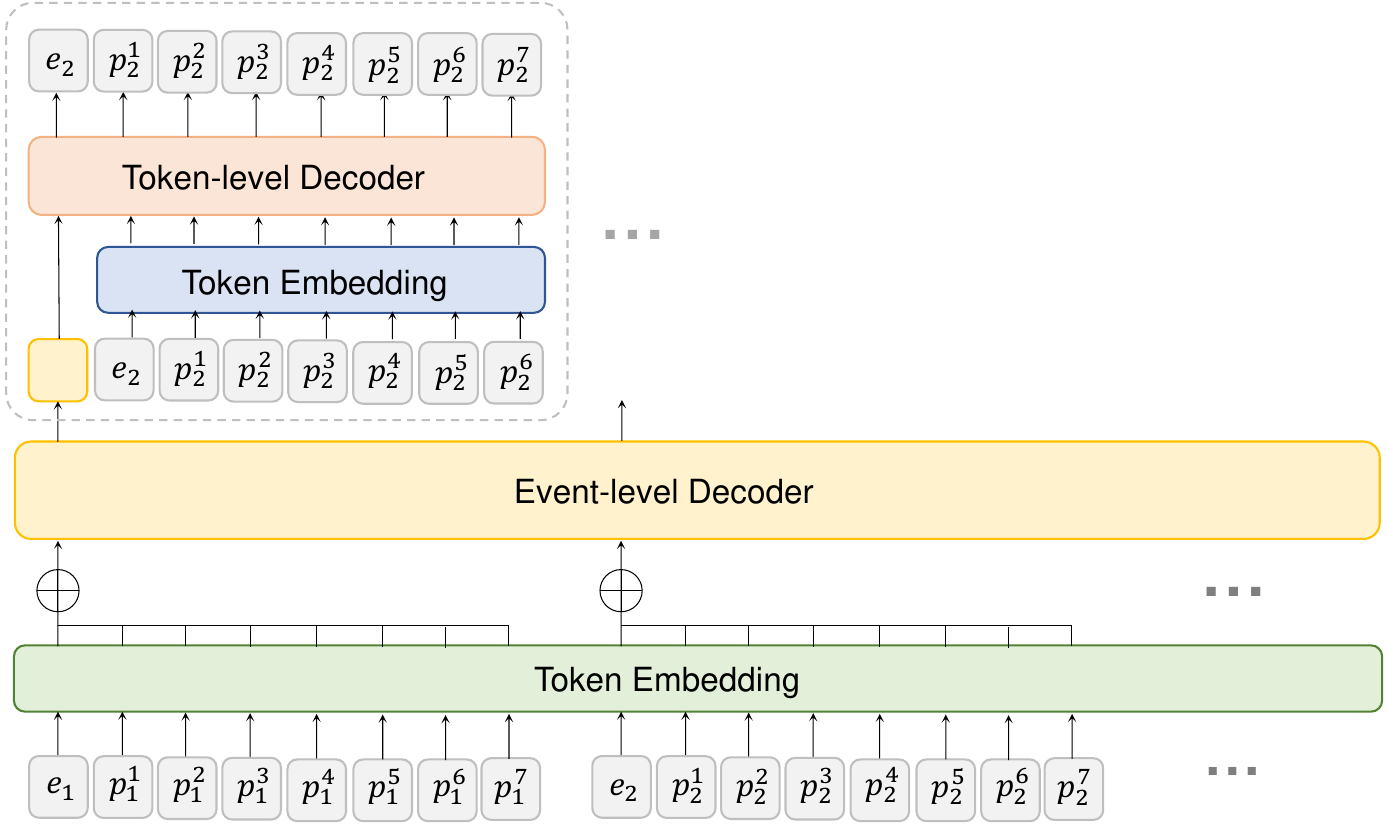} 
\caption{Illustration of the MIDI Event Transformer architecture, showcasing its two hierarchical decoders: the event-level decoder, which models temporal dependencies across high-level events, and the token-level decoder, which generates the detailed token sequence in an auto-regressive manner.}
\label{fig:example}
\end{figure}

\newpage

\section{Distribution of the Fine-tuning Dataset}

\begin{figure}[!b]
    \centering
    \begin{minipage}{0.41\textwidth}
        \centering
        \includegraphics[width=\textwidth]{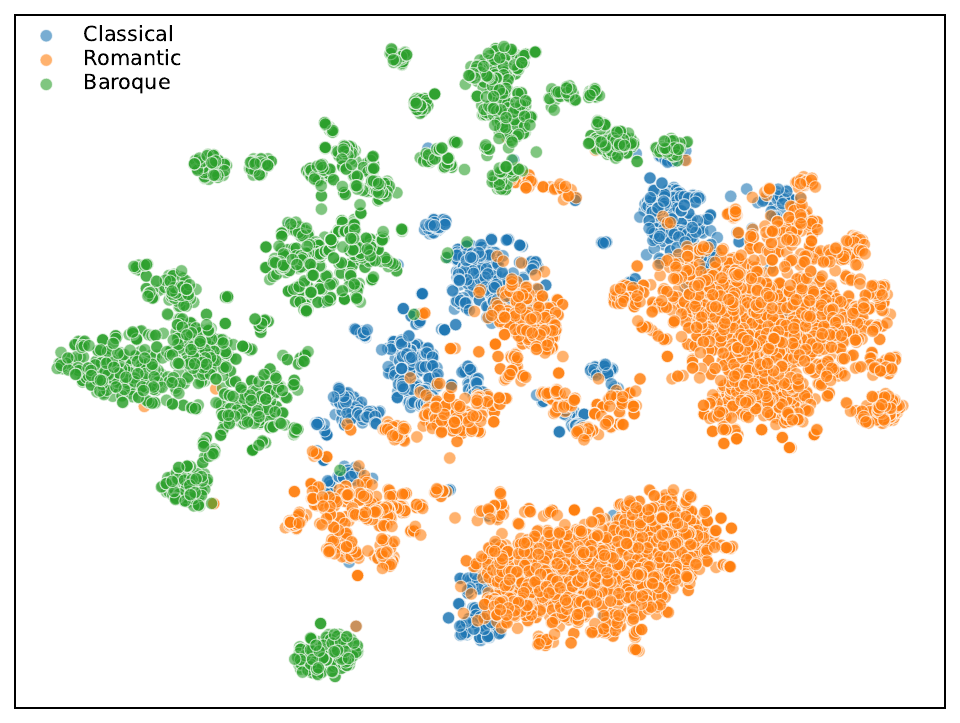}
        \caption*{(a) Distribution on periods.}
    \end{minipage} \\[0.5em]
    \begin{minipage}{0.41\textwidth}
        \centering
        \includegraphics[width=\textwidth]{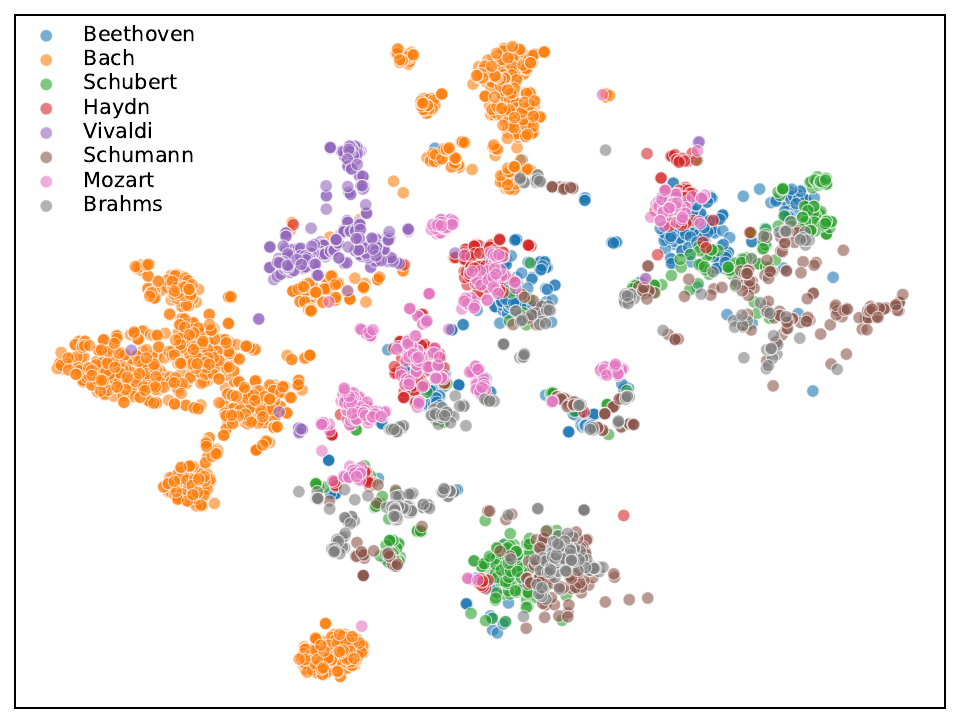}
        \caption*{(b) Distribution on most frequent composers.}
    \end{minipage} \\[0.5em]
    \begin{minipage}{0.41\textwidth}
        \centering
        \includegraphics[width=\textwidth]{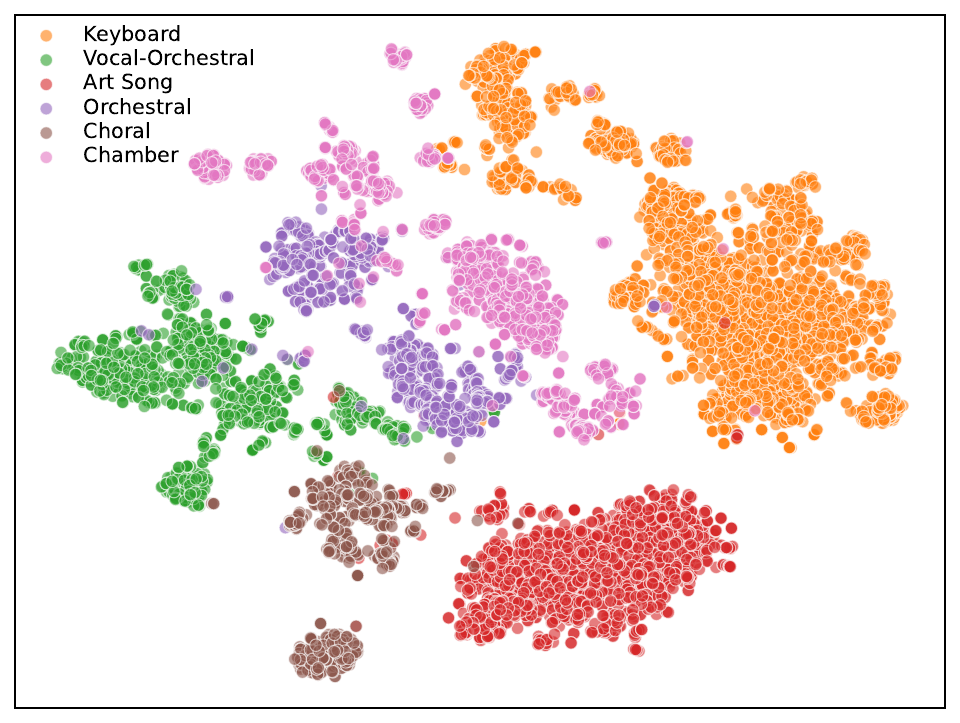}
        \caption*{(c) Distribution on instrumentations.}
    \end{minipage}
    \caption{\textit{t}-SNE visualizations for each label group on the fine-tuning dataset.}
    \label{fig:Figure 3}
\end{figure}

We utilized CLaMP 2 to extract the semantic features of all pieces, and the distribution of each label group is shown in Figure \ref{fig:Figure 3} (for composers, only eight composers with most pieces in the dataset are shown).

\section{Prompt Set in Experiments}

The full list of prompt set $P$ in the reinforcement learning stage of our experiments is listed in Table \ref{tab:promptset}.

\begin{table}[!htbp]
    \centering
    \begin{tabular}{@{}lll@{}}
        \toprule
        Period    & Composer                  & Instrumentation \\
        \midrule
        Baroque & Bach, Johann Sebastian & Chamber \\
        Baroque & Bach, Johann Sebastian & Choral \\
        Baroque & Bach, Johann Sebastian & Keyboard \\
        Baroque & Bach, Johann Sebastian & Orchestral \\
        Baroque & Bach, Johann Sebastian & Vocal-Orchestral \\
        Baroque & Corelli, Arcangelo & Chamber \\
        Baroque & Corelli, Arcangelo & Orchestral \\
        Baroque & Handel, George Frideric & Chamber \\
        Baroque & Handel, George Frideric & Keyboard \\
        Baroque & Handel, George Frideric & Orchestral \\
        Baroque & Handel, George Frideric & Vocal-Orchestral \\
        Baroque & Scarlatti, Domenico & Keyboard \\
        Baroque & Vivaldi, Antonio & Chamber \\
        Baroque & Vivaldi, Antonio & Orchestral \\
        Baroque & Vivaldi, Antonio & Vocal-Orchestral \\
        Classical & Beethoven, Ludwig van & Art Song \\
        Classical & Beethoven, Ludwig van & Chamber \\
        Classical & Beethoven, Ludwig van & Keyboard \\
        Classical & Beethoven, Ludwig van & Orchestral \\
        Classical & Haydn, Joseph & Chamber \\
        Classical & Haydn, Joseph & Keyboard \\
        Classical & Haydn, Joseph & Orchestral \\
        Classical & Haydn, Joseph & Vocal-Orchestral \\
        Classical & Mozart, Wolfgang Amadeus & Chamber \\
        Classical & Mozart, Wolfgang Amadeus & Choral \\
        Classical & Mozart, Wolfgang Amadeus & Keyboard \\
        Classical & Mozart, Wolfgang Amadeus & Orchestral \\
        Classical & Mozart, Wolfgang Amadeus & Vocal-Orchestral \\
        Classical & Paradis, Maria Theresia von & Art Song \\
        Classical & Reichardt, Louise & Art Song \\
        Classical & Saint-Georges, Joseph Bologne & Chamber \\
        Classical & Schroter, Corona & Art Song \\
        Romantic & Bartok, Bela & Keyboard \\
        Romantic & Berlioz, Hector & Choral \\
        Romantic & Bizet, Georges & Art Song \\
        Romantic & Boulanger, Lili & Art Song \\
        Romantic & Boulton, Harold & Art Song \\
        Romantic & Brahms, Johannes & Art Song \\
        Romantic & Brahms, Johannes & Chamber \\
        Romantic & Brahms, Johannes & Choral \\
        Romantic & Brahms, Johannes & Keyboard \\
        Romantic & Brahms, Johannes & Orchestral \\
        Romantic & Burgmuller, Friedrich & Keyboard \\
        Romantic & Butterworth, George & Art Song \\
        Romantic & Chaminade, Cecile & Art Song \\
        Romantic & Chausson, Ernest & Art Song \\
        Romantic & Chopin, Frederic & Art Song \\
        Romantic & Chopin, Frederic & Keyboard \\
        Romantic & Cornelius, Peter & Art Song \\
        Romantic & Debussy, Claude & Art Song \\
        Romantic & Debussy, Claude & Keyboard \\
        \bottomrule
        \multicolumn{3}{r}{Continued on next page}
    \end{tabular}
\end{table}

\begin{table}[!htbp]
    \centering
    \begin{tabular}{@{}lll@{}}
        \multicolumn{3}{r}{Table 2 -- Continued} \\
        \toprule
        Period    & Composer                  & Instrumentation \\
        \midrule
        Romantic & Dvorak, Antonin & Chamber \\
        Romantic & Dvorak, Antonin & Choral \\
        Romantic & Dvorak, Antonin & Keyboard \\
        Romantic & Dvorak, Antonin & Orchestral \\
        Romantic & Faisst, Clara & Art Song \\
        Romantic & Faure, Gabriel & Art Song \\
        Romantic & Faure, Gabriel & Chamber \\
        Romantic & Faure, Gabriel & Keyboard \\
        Romantic & Franz, Robert & Art Song \\
        Romantic & Gonzaga, Chiquinha & Art Song \\
        Romantic & Grandval, Clemence de & Art Song \\
        Romantic & Grieg, Edvard & Keyboard \\
        Romantic & Grieg, Edvard & Orchestral \\
        Romantic & Hensel, Fanny & Art Song \\
        Romantic & Holmes, Augusta Mary Anne & Art Song \\
        Romantic & Jaell, Marie & Art Song \\
        Romantic & Kinkel, Johanna & Art Song \\
        Romantic & Kralik, Mathilde & Art Song \\
        Romantic & Lang, Josephine & Art Song \\
        Romantic & Lehmann, Liza & Art Song \\
        Romantic & Liszt, Franz & Keyboard \\
        Romantic & Mayer, Emilie & Chamber \\
        Romantic & Medtner, Nikolay & Keyboard \\
        Romantic & Mendelssohn, Felix & Art Song \\
        Romantic & Mendelssohn, Felix & Chamber \\
        Romantic & Mendelssohn, Felix & Choral \\
        Romantic & Mendelssohn, Felix & Keyboard \\
        Romantic & Mendelssohn, Felix & Orchestral \\
        Romantic & Munktell, Helena & Art Song \\
        Romantic & Parratt, Walter & Choral \\
        Romantic & Prokofiev, Sergey & Keyboard \\
        Romantic & Rachmaninoff, Sergei & Choral \\
        Romantic & Rachmaninoff, Sergei & Keyboard \\
        Romantic & Ravel, Maurice & Art Song \\
        Romantic & Ravel, Maurice & Chamber \\
        Romantic & Ravel, Maurice & Keyboard \\
        Romantic & Saint-Saens, Camille & Chamber \\
        Romantic & Saint-Saens, Camille & Keyboard \\
        Romantic & Saint-Saens, Camille & Orchestral \\
        Romantic & Satie, Erik & Art Song \\
        Romantic & Satie, Erik & Keyboard \\
        Romantic & Schubert, Franz & Art Song \\
        Romantic & Schubert, Franz & Chamber \\
        Romantic & Schubert, Franz & Choral \\
        Romantic & Schubert, Franz & Keyboard \\
        Romantic & Schumann, Clara & Art Song \\
        Romantic & Schumann, Robert & Art Song \\
        Romantic & Schumann, Robert & Chamber \\
        Romantic & Schumann, Robert & Choral \\
        Romantic & Schumann, Robert & Keyboard \\
        Romantic & Scriabin, Aleksandr & Keyboard \\
        Romantic & Shostakovich, Dmitry & Chamber \\
        Romantic & Shostakovich, Dmitry & Keyboard \\
        Romantic & Sibelius, Jean & Keyboard \\
        Romantic & Smetana, Bedrich & Keyboard \\
        \bottomrule
        \multicolumn{3}{r}{Continued on next column}
    \end{tabular}
\end{table}

\begin{table}[!t]
    \centering
    \begin{tabular}{@{}lll@{}}
        \multicolumn{3}{r}{Table 2 -- Continued} \\
        \toprule
        Period    & Composer                  & Instrumentation \\
        \midrule        
        Romantic & Tchaikovsky, Pyotr & Keyboard \\
        Romantic & Tchaikovsky, Pyotr & Orchestral \\
        Romantic & Viardot, Pauline & Art Song \\
        Romantic & Warlock, Peter & Art Song \\
        Romantic & Wolf, Hugo & Art Song \\
        Romantic & Zumsteeg, Emilie & Art Song \\
        \bottomrule
    \end{tabular}
    \caption{The list of prompt set $P$ in reinforcement learning experiments.}
    \label{tab:promptset}
\end{table}

\end{document}